\RequirePackage{fix-cm}
\documentclass[smallextended]{svjour3}       
\smartqed  
\usepackage{graphicx}
\begin{document}

\title{Geodesic family of spherical instantons and cosmic quantum creation
}

\titlerunning{Spherical instantons and the cosmic quantum creation}        

\author{Ramon Lapiedra \and Juan Antonio Morales-Lladosa}


%
\institute{Ramon Lapiedra \and Juan Antonio Morales--Lladosa  \at Departament
d'Astronomia i Astrof\'{\i}sica, Universitat
de Val\`encia, \\ E-46100 Burjassot, Val\`encia, Spain. \\ Observatori Astron\`omic, Universitat de
Val\`encia, \\ E-46980 Paterna, Val\`encia, Spain.\\
Tel.: +34-96-3543066\\
Fax: +34-96-3543084\\
\email{ramon.lapiedra@uv.es}\\
\email{antonio.morales@uv.es}}

\date{Received: date / Accepted: date}

\maketitle

\begin{abstract}
The Einstein field equations for any spherically symmetric metric and a geodesic perfect fluid source are cast in a canonical simple form, both for Lorentzian metrics and for instantons. Both kinds of metrics are explicitly written for the  Lema{\^{\i}}tre-Tolman-Bondi family and for a general  $\Lambda$-Friedmann-Lema{\^{\i}}tre-Robertson-Walker universe. In the latter case  (including of course the instanton version) we study whether the probability of quantum creation of our Universe  vanishes or not. It is found, in accordance with previous results, that only the closed model can have a nonzero probability for quantum creation. To obtain this result, we resort to general assumptions, which are satisfied in the particular creation case considered by Vilenkin. On the other hand, Fomin and Tryon suggested that the energy of a quantically creatable universe should vanish. This is in accordance with the above result in which only the closed 
$\Lambda$FLRW model is quantically creatable while the open and flat models are not. That is so since it can be seen that this closed model has vanishing energy while the open model and the limiting flat case (suitably perturbed) have both infinite energy.

\end{abstract}

\keywords{Spherical symmetry \and LTB metrics \and Instanton metrics}
\PACS{04.20.Cv \and  98.80.Qc}


\section{Introduction: general considerations}
\label{intro}

In two seminal papers \cite{Vilenkin-1982,Vilenkin-1983}, Vilenkin estimated the probability of creating from ``nothing'' a closed inflationary universe, that is, a closed de Sitter universe. Here, ``nothing'' is a state with, in particular, no time. In the present paper,  we revisit this issue and consider an Euclidean differentiable manifold, that is, one with signature $+ 4$, where there is also no time.

The solutions of the Einstein field equations for signature $+ 4$ are sometimes called Euclidean solutions or instantons and we will use this naming in the present paper. For Vilenkin, the creation mechanism is a quantum tunneling event to the classical closed de Sitter universe from the corresponding instanton, whose semiclassical probability, $P$, is estimated as $P \propto \exp{(-|S_{E}|)}$ (see \cite{Vilenkin-1984,Vilenkin-1985};  and also \cite{Zeldovich-Starobinsky-84}),  where the pre--exponential factor has been omitted, and $S_{E}$ ($S_{E} / \hbar$, if we do not take $\hbar = 1$) is the corresponding instanton action.

Actually, the semiclassical expression for $P$, that we will adopt throughout the present paper, is not tied to some tunneling event but to the quantum probability amplitude of going from an ``initial'' wave function of a universe to a ``final'' one, in the semiclassical approximation of the path integral version of quantum gravity \cite{HarHaw-1983}. Here, ``initial'' and ``final'' do not refer to time, now (in  an Euclidean manifold) inexistent, let it be external or not \cite{Zeh-1988}. These terms refer to a vanishing value, $u = 0$, and a turning point value, $u = u_m$, of the new spatial coordinate, $u$, of the corresponding instanton, that replaces the time of the considered Lorentzian solutions of the Einstein equations (see next subsection \ref{sec:5c}). In accordance with  \cite{HarHaw-1983},  the approximate expression for $P$ will be
 \begin{equation}\label{proba}
 P \propto \exp{(-S_{E})}. 
 \end{equation}

In the current literature (se also, for instance, \cite{Linde-1984a,Linde-1984b}, besides the above references) there is, nevertheless, some vacillations as far as the sign of the exponent in (\ref{proba}) is concerned. A discussion of this point can be found, for example, in \cite{Vilenkin-1999}, or in the book \cite{Kolb-Turner-1990}.

On the other hand, in the particular case considered in  \cite{Vilenkin-1983}, the situation is simple enough to find the instanton corresponding to the closed de Sitter universe by simple changing $\dot{a}^2 \equiv (\partial_t a)^2$, with $t$ the cosmic time and $a(t)$ the cosmic expansion factor, by $-\dot{a}^2$ (i.e. changing $t$ by $i \, t$). 
However in more complex cases, such as the one considered in \cite{AtkaPag-82} (where it was shown that the only creatable FLRW model is the closed one) and 
\cite{Dabrowski-Larsen-95},  the corresponding Einstein field equations must be integrated in both, the Euclidean (signature $+ 4$) and the Lorentzian (signature $+ 2$) cases. That is, in these more complex cases, as we do in the present paper, we should perform this double integration and then choose the specific instanton from which we are planning to obtain by quantum creation the selected model of our present universe.

In the present paper we will consider the quantum creation of the general  $\Lambda$-Friedmann-Lema{\^{\i}}tre-Robertson-Walker 
($\Lambda$FLRW) universes, with $\Lambda$ the cosmological constant. More specifically we will roughly estimate, in the semiclassical approximation, the quantum probability of going to one of these $\Lambda$FLRW universes from the corresponding instanton, starting from the general path integral way developed in \cite{HarHaw-1983} for quantum gravity. To this end, extending previous work by Ellis \cite{Ellis-1992} and Hellaby et al. \cite{HeSuEllis-1997}, we will first give a simple canonical form for the Einstein field equations with spherical symmetry when the source is a geodesic perfect fluid both for the Lorentzian  and the Euclidean (instanton) metrics. Furthermore, we will recover the explicit solutions (Lorentzian and instanton) of these equations in the particular case  of the Lema{\^{\i}}tre-Tolman-Bondi (LTB) metric family. This approach  includes the 
$\Lambda$FLRW universes  as a particular case, and allows us to recover previous results concerning signature changes in FLRW cosmologies 
\cite{Ellis-1992,HeSuEllis-1997}. 

In our study, we will assume a general criterion to write the expression of the ``stress-energy'' tensor in the instanton case. Finally we advance the criterion to select, given a classical metric, the particular instanton from which this classical metric could be perhaps created. These criteria, that will be stated next more specifically (see section \ref{sec:4}), are all them fulfilled in the particular case considered in \cite{Vilenkin-1983}. 

The present paper goes along the following lines. In Sect. \ref{sec:2}, following the procedure outlined in  \cite{Ellis-1992,HeSuEllis-1997,Ellis-Su-Co-He-1992}, in order to include an isotropic pressure and a cosmological constant, the Einstein field equations are reduced to a  canonical simple form, both in the Lorentzian and in the Euclidean case. This is done for a family of metrics with spherical symmetry, whose energy content in the Lorentzian case is a geodesic perfect fluid, with homogeneous pressure, inhomogeneous density and cosmological constant. In particular, we give this double general solution (Lorentzian and Euclidean) of the LTB family of metrics, that is, when pressure and cosmological constant vanish. Technical details are provided in Appendix \ref{ap-B}.  In Sect. \ref{sec:3} we revisit  the particular case of the general $\Lambda$FLRW metrics and their instanton counterparts, governed by the equations stated in \cite{Ellis-1992} for $\Lambda = 0$. In Sect. \ref{sec:4} we discuss the general criteria we use to establish the possibility of quantum creation for our universe. We verify that these criteria are fulfilled in the particular case treated in \cite{Vilenkin-1983}. In Sect. \ref{sec:5} we study the possibility of creation of the closed model of our present universe. In accordance with the results in \cite{AtkaPag-82}, and also with the considerations in \cite{Vilenkin-1999},  but following a different method, we conclude that the closed general $\Lambda$FLRW universe has a finite semiclassical probability of being quantically created. We comment on the differences of our procedure and the approaches considered by other authors \cite{Vilenkin-1999,AtkaPag-82}. Finally, our results are discussed in Sect. \ref{sec:6}, where we conclude briefly that the probability of quantum creation for an open non flat or a flat $\Lambda$FLRW model vanishes (in accordance with \cite{AtkaPag-82}, too).

For the sake of completeness, we add three appendices where some calculations are presented in detail. The results presented in these appendices concern the inclusion of the Lorentzian and Euclidean cases in a common treatment and are  explicitly stated in the references quoted there or easily deduced from them.

A short report containing some results, without proof,  of this
work recently appeared in the Proceedings of the Spanish Relativity Meeting
ERE-2014 \cite{ere2014}.
%
%
%


\section{A large family of instantons}
\label{sec:2}

Following previous works  \cite{Ellis-1992,HeSuEllis-1997,Ellis-Su-Co-He-1992}, let us consider the Einstein field equations with cosmological constant $\Lambda$, that is, 
\begin{equation}\label{EEsignatura}
R_{\alpha\beta} - (\frac{R}{2}- \Lambda) g_{\alpha\beta} = \kappa T_{\alpha\beta}  \quad (\kappa = 8 \pi G, \, c=1), 
\end{equation}
both for Lorentzian and Euclidean metrics, in the case of spherical symmetry, with inhomogeneous stress-energy tensor corresponding to a perfect fluid. This stress-energy tensor, $T^{\alpha \beta}$, reads out 
\begin{equation}\label{Tmunu}
T^{\alpha \beta} = \bar{p} g^{\alpha \beta} + (\bar{\mu} + \bar{p}) n^\alpha n^\beta, \quad g_{\alpha \beta} n^\alpha n^\beta = \epsilon,
\end{equation}
with $\bar{\mu}$ the energy density, $\bar{p}$ the pressure and $\epsilon = \pm 1$, the $+1$ value corresponding to the Euclidean solution and 
$\epsilon = -1$ to the Lorentzian one. For later convenience we will write 
\begin{equation}\label{Tmunu2}
T^{\alpha \beta} = \mu n^\alpha n^\beta + p \, \gamma^{\alpha\beta}, \quad \gamma^{\alpha \beta}= g^{\alpha \beta} - \epsilon \, n^\alpha n^\beta, 
\end{equation}
the relation between $(\mu, p)$ and $(\bar{\mu}, \bar{p})$ becoming 
\begin{equation}\label{mupebar}
\mu = \bar{\mu} + (1 + \epsilon) \bar{p},  \quad p = \bar{p},
\end{equation}
such that for $\epsilon = -1$, $\mu = \bar{\mu}$.

These Einstein field equations, both for the Lorentzian and the Euclidean case,  can be reduced to the following differential equations (see Appendix \ref{ap-B}, Eqs. (\ref{A equation}) and  (\ref{A equation 2})):
\begin{equation}\label{flu-insta1}
\dot{A}^2+2A\ddot{A} + K + \epsilon \, \Lambda \, A^2 = \epsilon \, \kappa \, p  \, A^2, 
\end{equation}
and
\begin{equation}\label{flu-insta2}
2 \, \frac{\ddot{A}}{A}+\frac{\ddot{A'}}{A'} + \, \epsilon \,  \Lambda =- \frac{\kappa}{2} \, (\mu  -3 \, \epsilon \, p),     
\end{equation}
in Gauss coordinates $(u, \rho, \theta, \phi)$ adapted  to the spherical symmetry such that  $n^\alpha=(1,0,0,0)$, provided that $p$ only depends on $u$ and $A'$ is not identically zero.%
%
\footnote{In the case $A'=0$, the integration of the corresponding Einstein equations leads to a generalization of the well  known Datt solution (see \cite{PlebKra} and related references quoted therein) and its corresponding instanton.  \label{sobre-Datt}}
%
%
%
 In these coordinates the line element becomes (see Appendix \ref{ap-B}, Eq. (\ref{A-LTBmetric})):
\begin{equation}\label{LTB-inst}
ds^2= \epsilon \,  du^2+\frac{A'^2(u,\rho)}{1+ \epsilon \, K(\rho)} \, d\rho^2+A^2d\sigma^2, \, A' \equiv \partial_\rho A 
\end{equation}
with $K(\rho)$, $K$ for brevity,  an arbitrary function of $\rho$,  $d\sigma^2$ the metric of the unit two-sphere, and the dot indicating derivation with respect to $u$ and the prime derivation with respect to $\rho$.

Here, when we put $\epsilon = -1$ (the Lorentzian case), we can put $u = t$ for the corresponding time. In this case, the family of metrics considered corresponds to a generalization of the well known LTB metrics, the particular family that we obtain by putting $p = \Lambda = 0$, for the pressure  and the cosmological constant, respectively.

The general solution of Eqs. (\ref{flu-insta1}), (\ref{flu-insta2}), in the LTB case \cite{Lemaitre-1933,Tolman-1934,Bondi-1947}, is well known and is expressed 
as \cite{Landau}:
\begin{equation} \label{kp}
{\rm If} \, K(\rho) > 0,  \, \left \{
\begin{array} {ll}
A(u, \rho) = \displaystyle{\frac{M(\rho)}{K(\rho)}(1 - \cos \eta)}, \\ \\
\eta - \sin\eta = \displaystyle{\frac{[K(\rho)]^{3/2}}{M(\rho)} [u - \psi(\rho)]}.
\end{array} \right.
\end{equation}
\begin{equation} \label{k0}
{\rm If} \, K(\rho) = 0, \,  A(u, \rho) = \displaystyle{\Big\{\frac{9}{2} M(\rho) [u - \psi(\rho)]^2\Big\}^{1/3}}.
\end{equation}
\begin{equation} \label{kn}
{\rm If} \, K(\rho) < 0, \, \left \{
\begin{array} {ll}
A(u, \rho) = \displaystyle{ \frac{M(\rho)}{K(\rho)}(1 - \cosh \eta)},\\ \\
\sinh\eta - \eta = \displaystyle{\frac{[- K(\rho)]^{3/2}}{M(\rho)} [u - \psi(\rho)]}.
\end{array} \right.
\end{equation}
where $M(\rho)$ and  $\psi(\rho)$ are arbitrary functions of $\rho$, and $\eta$ is a suitable parameter. 

The main remarkable consequence of the presentation followed in this section is that the $\epsilon$ sign does not appear in the above expressions of the general solution. It only appears explicitly in the line element (\ref{LTB-inst}). Function $K(\rho)$ is related to the geometry of the  
3-surfaces $t = constant$. In fact, these 3-surfaces are flat if, and only if, $K(\rho)$ is identically zero. In the Lorentzian case, $M(\rho)$ gives the ``effective'' gravitational mass of the dust configuration which coincides with the total mass only when $K(\rho)$ vanishes; $ \psi(\rho)$ gives the proper time when the essential singularity $A(t, \rho) =0$ occurs (for further considerations, see \cite{PlebKra}).


\section {General $\Lambda$FLRW metrics and their instanton counterparts} 
\label{sec:3}

Let us consider the particular solution of (\ref{flu-insta1}) and  (\ref{flu-insta2}), where the $A(u, \rho)$ function factorizes, that is
\begin{equation}\label{factoritza}
A(u, \rho) = a(u) \, f(\rho)
\end{equation}
where $a(u)$ and $f(\rho)$ are two functions to be determined. Furthermore, in (\ref{flu-insta2}) besides having $p = p(u)$ we will put
\begin{equation}\label{mupeu}
\mu = \mu(u), 
\end{equation}
that is, besides $p$, $\mu$ does not depend on $\rho$.

Substituting (\ref{factoritza}) and (\ref{mupeu}) in (\ref{flu-insta1}) and (\ref{flu-insta2}), having in mind that $p =p(u)$, we obtain in full agreement with \cite{Ellis-1992}:
\begin{equation}\label{flu-insta1F}
\frac{\dot{a}^2}{a^2} = \frac{\kappa}{3} \, \mu - \frac{k}{a^2} - \frac{\epsilon}{3} \,  \Lambda, 
\end{equation}
\begin{equation}\label{flu-insta2F}
\frac{\ddot{a}}{a} = - \frac{\kappa}{6} \,  (\mu - 3 \,  \epsilon \, p) - \frac{\epsilon}{3} \,  \Lambda, 
\end{equation}
that imply the ``energy conservation'' equation
\begin{equation}\label{sobre-n-orto-gen}
\dot{\mu} = - 3 (\mu - \epsilon p) \, \frac{\dot{a}}{a}\, .
\end{equation}
More specifically it follows from Eq. (\ref{energy-inst}) in the Appendix \ref{ap-A} by taking $\theta \equiv \nabla_\alpha n^\alpha = 3 \dot{a}/a$ for the $\Lambda$FLRW case.  

For $\epsilon = -1$, Eqs.  (\ref{flu-insta1F}) and  (\ref{flu-insta2F}) reproduce the cosmic dynamical equations for the expansion factor, $a(t)$, of a $\Lambda$FLRW universe, $k = +1, 0, -1$ being the normalized constant with the sign of $K(\rho)$ (curvature index). Further, for the function $K$ in (\ref{LTB-inst}) we obtain $K = k \rho^2$ if we take $A^2 = a^2 \rho^2$.

Finally, in order to write the corresponding element of line, $ds^2$, we only have to put this expression of $A^2$, with the corresponding $a$ solution of (\ref{flu-insta1F}), (\ref{flu-insta2F}), in Eq. (\ref{LTB-inst}). We obtain%
%
\footnote{Notice that, according to (\ref{LTB-instF}), an instanton with $k=-1$ is a closed space, i.e., $\rho \in (0,1)$. Actually, since for an instanton it is $\epsilon = +1$, the function $1+\epsilon k \rho^2$ in  (\ref{LTB-instF}) becames then $1 - \rho^2$. Similarly, an instanton with $k = +1$ is an open space: $1+\epsilon k \rho^2$ becomes $1 + \rho^2$   
and $\rho \in (0, + \infty)$. \label{sobre-it}}
%
%
%
\begin{equation}\label{LTB-instF}
ds^2= \epsilon \, du^2+ a^2 \Big( \frac{ d\rho^2}{1+ \epsilon \, k \rho^2} + \rho^2 d\sigma^2\Big). 
\end{equation}

In particular, for $\epsilon = +1$ the instanton metric is 
\begin{equation}\label{instFLRW}
ds^2_E=  du^2+ a^2_E \Big( \frac{ d\rho^2}{1+ \, k \rho^2} + \rho^2 d\sigma^2\Big),  
\end{equation}
with $a_E \equiv a(u)$ a solution of  (\ref{flu-insta1F}) and (\ref{flu-insta2F}) with $\epsilon = + 1$. That is, we have for $a(u)$ the equations:
\begin{equation}\label{flu-insta1F+}
\frac{\dot{a}^2}{a^2}  =  \frac{\kappa}{3} \, \mu  - \frac{k}{a^2} - \frac{\Lambda}{3}, 
\end{equation}
\begin{equation}\label{flu-insta2F+}
\frac{\ddot{a}}{a} =  - \frac{\kappa}{6} \, (\mu - 3\, p ) - \frac{\Lambda}{3}.
\end{equation}

Leaving out the mere inclusion of the cosmological constant $\Lambda$, the results of this section are included in those 
of the reference \cite{Ellis-1992} by Ellis.


\section{Calculating the suitable instantons for any $\Lambda$FLRW universe.} 
\label{sec:4}

As we have commented in the Introduction, in the present paper we want to estimate the quantum probability of going from some suitable Euclidean solution (instanton) to some particular universe. Then, we need to calculate these instantons, a calculation that we perform in subsections \ref{sec:4b} and \ref{sec:4c} for a family of different cases. Let us be more specific:
\begin{enumerate}
\item[(a)] First, by definition,  the instantons satisfy the Einstein field equations for a metric of signature $+4$, whose ``stress-energy'' tensor is {\em formally} the same as the Lorentzian one, ``formally'' meaning that the only change introduced in it when jumping from $\epsilon = -1$ to $\epsilon = +1$ is to change the original Lorentzian metric, let us say, $g_{\alpha\beta} (\epsilon = -1)$, 
$\alpha, \beta = 0, 1, 2, 3$, by the new Euclidean one $g_{\alpha\beta} (\epsilon = +1)$ (see Eqs. (\ref{Tmunu2}) and (\ref{LTB-instF})).  
\end{enumerate}

Also, in the $\Lambda$FLRW case, in order to associate, to each particular $\Lambda$FLRW metric, the corresponding instanton from which to test if that metric can become quantically created  we assume that: 
\begin{enumerate}
\item[(b)] The $\Lambda$FLRW metric and its corresponding instanton are suitably matched across the discontinuity 3-surface $u=t=0$ of the $\Lambda$FLRW  metric, thus extending to a singular 3-surface the matching prescription in \cite{Ellis-1992,HeSuEllis-1997,Ellis-Su-Co-He-1992}. These extended matching conditions are analogous to the Darmois continuity conditions, now for the differences of the first, $g_{ij}$,  and second, $K_{ij}$, fundamental forms through $u = t=0$. It is important to remark that  we need to speak 
of differences, that is,  of $g_{ij}-(g_E)_{ij}$ and $K_{ij}-(K_E)_{ij}$, where the $E$ subindex stands for ``Euclidean''. This is so since, at $t=0$, $K_{ij}$ could be now singular. However, in our extension of the matching Darmois conditions to a singular 3-surface,  the differences are assumed to be regular and vanishing (see next), in other words, to $u\to 0$, because of the kind of matching performed, $(K_E)_{ij}$ becomes singular at the same rate that $K_{ij}$, if $K_{ij}$ does.
\end{enumerate}

In dealing with a change of signature on a space-like non-singular 3-surface, the fulfillment of the Darmois matching conditions is not the sole alternative that has been proposed in the past (cf \cite{Hellaby-Dray-1994,Hayward-1995,Hellaby-Dray-1995}). In fact, according to \cite{Hayward-1995,Gibbons-Hartle-1990},  such a change might imply the vanishing of $K_{ij}$  on the junction surface. An accurate analysis of the behaviour of the different (scalar, vector and tensor) types of cosmological perturbations under this more restrictive matching condition ($K_{ij} = 0$ on the signature change 3-surface) has been carried out in \cite{JMartin-1995}; but this question is going beyond the scope of the present paper.

As we will show next in detail, the particular tunneling creation considered in \cite{Vilenkin-1983} fulfills these two demands on the junction surface, $[K_{ij}]=0$ 
(see the second equation in Eq. (\ref{disco}) below) and $K_{ij} = 0$, as it can be easily deduced from Eqs. (\ref{a-int-dS-knegativa}) and (\ref{deSitter-solucioV}). The opposite situation happens for any other $\Lambda$FLRW universe, as we explain at the end of the subsection \ref{sec:4c}.

At this point, the motivation to introduce assumption (b) would seem perhaps insufficiently justified both from physical and mathematical points of view. In fact, on a singular 3-surface some algebraic invariant of the curvature tensor diverges. Nevertheless, according to the spirit of the assumption (b),  it should be sufficient to require that the corresponding Lorentzian and Euclidean curvature invariants diverge at the same ratio in reaching the signature change 3-surface. In the case of the generalized LTB model, this condition is fulfilled both for the the four dimensional Ricci curvature and the Kretschmann scalars (for the respective expressions, see Eqs. (48) and (49) in reference \cite{HeSuEllis-1997}). In addition, we are going to show that our assumption (b) is naturally consistent with the paradigmatic physical situation concerning the quantum creation of a closed de Sitter universe and the involved mathematical treatments (see subsection \ref{sec:4b}).

%


\subsection{Matching conditions under signature change} 
\label{sec:4a}

Let us be more specific when stating the above (b) assumption, the ``extended Darmois matching conditions''. These conditions assume the vanishing of the three-space metric difference $g_{ij}-(g_E)_{ij}$ (in our case given implicitly in (\ref{LTB-instF})) and their corresponding extrinsic curvature diference, $K_{ij}-(K_E)_{ij}$, through the matching 3-surface, $u = t = 0$ in the present case,  in which case  $K_{ij}$ becomes $K_{ij} = - \frac{1}{2} \partial_u g_{ij}$. Notice that according to (\ref{LTB-instF}), the vanishing of  
$g_{ij}-(g_E)_{ij}$ and $K_{ij}-(K_E)_{ij}$,  through $u = t = 0$, that we will write
\begin{equation}\label{disco}
[g_{ij}] = 0,  \quad [K_{ij}] = 0, 
\end{equation}
reduces --using the same notation-- to: 
\begin{equation}\label{disco-a}
[a] = 0,  \quad [\dot{a}] = 0, 
\end{equation}
that is (denoting $y(x)|_{x=0} \equiv \lim_{x \to 0} y(x)$)
\begin{equation}\label{matching}
a(t)|_{t=0} - a_{E}(u)|_{u=0}= 0, \,   \dot{a}(t)|_{t=0} - \dot{a}_{E}(u)|_{u=0} = 0, 
\end{equation}
and 
\begin{equation}\label{matching-grho}
g_{\rho\rho}(t)|_{t=0} = g_{\rho\rho}(u)|_{u=0},  
\end{equation}
which  imply that  the sign $\epsilon k$ in (\ref{LTB-instF}) does not change across the matching 3-surface $t=u=0$.%
%
\footnote{Notice that, both in the closed and open FLRW models, $\dot{a}(t)|_{t=0}$  unavoidably  diverges because of the physical singularity present in $t=u=0$. Nevertheless, as stated above, the corresponding extended Darmois condition is written as  $\dot{a}(t)|_{t=0} - \dot{a}_{E}(u)|_{u=0} = 0$, indicating that, for $u=0$,  
$\dot{a}_{E}(u)|_{u=0}$ diverges in the suitable way to make this difference vanish.}%
%
\ Consequently, 
\begin{equation}\label{epsilonk}
(\epsilon k)_E = (\epsilon k)_L, 
\end{equation}
where the subscripts $E$ and $L$ stand for ``Euclidean'' and ``Lorentzian'', respectively. Then, from Eq. (\ref{flu-insta1F}),  
\begin{equation}\label{matching2}
\Big(\frac{\kappa}{3} \, \mu - \frac{k_L}{a^2} + \frac{\Lambda}{3} \Big)_{t=0} =  \Big(\frac{\kappa}{3} \, \mu_E - \frac{k_E}{a^2} - \frac{\Lambda}{3}\Big)_{u=0}, 
\end{equation}
with, because of (\ref{epsilonk}), $k_E = - k_L$.

Then, if we require that the instanton solution exists  for a bounded $\rho$ coordinate domain,  we must take $k_E = -1$, and so $k_L = +1$. 
In this way, we will have matched the closed instanton solution
\begin{equation}\label{instFLRWmenys}
ds^2_E=   du^2+ a^2_E(u)\Big( \frac{ d\rho^2}{1- \rho^2} + \rho^2 d\sigma^2\Big), 
\end{equation}
and the closed FRLW metric
\begin{equation}\label{closedFLRW}
ds^2= - dt^2+ a^2(t) \Big( \frac{ d\rho^2}{1- \rho^2} + \rho^2 d\sigma^2\Big),  
\end{equation}
across the 3-surface $u=t=0$, once we apply the condition (\ref{disco-a}) to the above $a_E$ and $a$ functions, which must satisfy  the corresponding Einstein equations 
(\ref{flu-insta1F}) and (\ref{flu-insta2F}), with $\epsilon = +1$ for $a_E$, or $\epsilon = -1$ for $a$.


\subsection{The particular case of a closed de Sitter universe} 
\label{sec:4b}

Before continuing, let us consider the particular case considered by Vilenkin \cite{Vilenkin-1983}, corresponding to $\mu = p = 0$ and  $\Lambda > 0$. In this case we never can  have  $k = +1$ in (\ref{flu-insta1F+}). Thus Eqs. (\ref{flu-insta1F+}) and (\ref{flu-insta2F+}) become
\begin{equation}\label{flu-insta1deS}
\dot{a}^2 =  1 - \frac{\Lambda}{3} \, a^2,  \qquad \ddot{a}=  - \frac{\Lambda}{3} \, a,
\end{equation}
with $a$ standing for $a(u) = a_E$.

A particular instanton solution of these equations is the one calculated in \cite{Vilenkin-1983} 
\begin{equation}\label{a-int-dS-knegativa}
a(u) = \sqrt{\frac{3}{\Lambda}} \, \cos \Big(\sqrt{\frac{\Lambda}{3}} \, u\, \Big), 
\end{equation}
to be put in the metric (\ref{instFLRWmenys}).  Further,  changing $\rho$ to $\chi$ by $\rho = \sin \chi$, (\ref{instFLRWmenys}) becomes
\begin{equation}\label{inst-deSitter-tancat-chi}
ds^2_E=   du^2+ a^2_E(u) ( d \chi^2 + \sin^2 \chi \,  d\sigma^2). 
\end{equation}

We see that $u$ and $\chi$ take both values in a finite domain, in particular we take $u \in [0, \frac{\pi}{2}\sqrt{\frac{3}{\Lambda}}]$ and  $\chi \in [0, 2 \pi]$. 
Because of this fact Vilenkin \cite{Vilenkin-1985} obtains a finite value for the corresponding instanton action $S_{E}$ 
(see Eq. (\ref{accioE2}) below). In our notation this finite value is 
\begin{equation}\label{Inst-accio-deS}
 S_{E} =  \frac{3 \pi}{G \Lambda}, 
\end{equation}
that gives a finite probability of creation, but of which one de Sitter universe:%
%
\footnote{No privileged family of cosmic observers exists in the de Sitter space-time (because it is ``empty'') but there exist three different families of Gaussian 
observers admitting orthogonal hypersurface of constant curvature, and then three different metric forms (closed, flat or open) for the three-space line element expressed 
in the respective comoving coordinates. Of course, there also exist static (non Gaussian) observers having associated the 
de Sitter static metric form (see, for instance, \cite{Robertson-Noonan}). Here, we are considering the {\em closed de Sitter universe}, that is,  the  metric form that result by adapting coordinates to a a family of Gaussian observers admitting orthogonal three spaces of constant positive curvature.
\label{de Sitter}}%
 %
\ the closed or the open one? The closed one, since above the use of the (b) matching conditions 
has led us to match (\ref{closedFLRW}) with (\ref{instFLRWmenys}). Thus,  the de Sitter created model ($\epsilon = -1$) is the closed one ($k = +1$) that is 
\begin{equation}\label{deSitter-tancat-chi}
ds^2 =   - dt^2+ a^2(t) ( d \chi^2 + \sin^2 \chi \,  d\sigma^2),  
\end{equation}
with $a(t)$ a solution of the Eqs. (\ref{flu-insta1F}) and (\ref{flu-insta2F}) for $\epsilon = -1$, $k=+1$, $\mu = p = 0$, that is to say, a solution of 
\begin{equation}\label{deSitter-dos}
\dot{a}^2 = -1 +  \frac{\Lambda}{3} \, a^2,  \qquad \ddot{a}=  \frac{\Lambda}{3} \, a,
\end{equation}
that we take
\begin{equation}\label{deSitter-solucioV}
a(t) = \sqrt{\frac{3}{\Lambda}} \, \cosh \Big(\sqrt{\frac{\Lambda}{3}} \,  t\, \Big).
\end{equation}

Eqs. (\ref{a-int-dS-knegativa}) and (\ref{deSitter-solucioV}) are also the choice made in \cite{Vilenkin-1983}, but, as we have just announced, in our case, this choice come from a general assumption (b), i.e., from the extended Darmois conditions as we have seen (conditions (\ref{disco-a}) and (\ref{epsilonk}) in the present case). 

The conditions are then satisfied by  (\ref{inst-deSitter-tancat-chi}) and  (\ref{deSitter-tancat-chi}). In all, we have proved that the choices made by Vilenkin \cite{Vilenkin-1983} when creating a closed de Sitter universe by quantum tunneling from a given instanton can be deduced  from our  two general assumptions (a) and (b), that in the next sections we will apply in order to test the quantum creatability of the rest of the different $\Lambda$FLRW universe models. This result about the de Sitter universe was previously obtained  by Ellis {\em et al.} \cite{Ellis-Su-Co-He-1992},  remarking the interest of the result in the  development of the Hartle-Hawking  quantum cosmology program \cite{HarHaw-1983}.
%


\subsection{The instanton limiting behaviour for a closed and very ``hot'' $\Lambda$FLRW universe}
\label{sec:4c}

Now, let us consider the case of a closed $\Lambda$FLRW universe largely dominated by radiation in an epoch prior to inflation. Setting $p = \mu/3$, the integration of (\ref{sobre-n-orto-gen}) when $\epsilon =-1$ gives  
\begin{equation}\label{murad}
\mu = \mu_e \, \Big(\frac{a_e}{a}\Big)^4, 
\end{equation}
index $e$ referring to some instant prior to the energy-matter equivalence epoch. Thus, for $t$ small enough,  the cosmic dynamical Eqs. (\ref{flu-insta1F}) and (\ref{flu-insta2F}), for the Lorentzian solution, $\epsilon = -1$,  and $p = \mu/3$, reduces to
\begin{equation}\label{2eqs-aprox-L}
\Big(\frac{\dot{a}}{a}\Big)^2 \simeq \frac{\kappa}{3} \, \mu,  \quad \quad \frac{\ddot{a}}{a} \simeq - \frac{\kappa}{3} \, \mu, 
\end{equation}
whose well known solution, by considering (\ref{murad}), becomes 
\begin{equation}\label{aLaprox}
a\simeq a_e \sqrt{\frac{t}{t_e}}, \quad t_e \equiv \frac{1}{2}\sqrt{\frac{3}{\kappa \mu_e}}. 
\end{equation}

On the other hand, taking into account (\ref{matching}) and  the first equation of (\ref{2eqs-aprox-L}), the condition (\ref{matching2}) becomes now
\begin{equation}\label{matching-mu}
\mu_E|_{u=0} = \mu|_{t=0}, 
\end{equation}
and then in the limit $u \to 0$,  for the instanton solution, $\epsilon = +1$, because of the $\mu$ term dominance, Eqs. (\ref{flu-insta1F+}) and (\ref{flu-insta2F+}) become,
\begin{equation}\label{2eqs-aprox}
\Big(\frac{\dot{a}_E}{a_E}\Big)^2 \simeq \frac{\kappa}{3} \, \mu, \quad \frac{\ddot{a}_E}{a_E} \simeq - \frac{\kappa}{6} \, (\mu - 3 p_E). 
\end{equation}
with  the corresponding integrability condition (\ref{sobre-n-orto-gen}) that now is
\begin{equation}\label{sobre-n-orto-E}
\dot{\mu}_E = - 3 (\mu_E - p_E) \, \frac{\dot{a}_E}{a_E}.
\end{equation}

Then, from (\ref{matching-mu})-(\ref{sobre-n-orto-E}), we show straight away  that the instanton solution for $u \to 0$ behaves as the Lorentzian solution (\ref{aLaprox}) for $t \to 0$. In fact, from (\ref{sobre-n-orto-E}), for $\mu$ dominated by radiation, having in mind (\ref{matching-mu}), we obtain $p_E = - \mu_E/3 = -\mu/3$ for $u \to 0$. Thus, in this limit, the second equation in (\ref{2eqs-aprox}) becomes:
\begin{equation}\label{a-dos-punts-approx}
 \frac{\ddot{a}_E}{a_E} \simeq - \frac{\kappa}{3} \, \mu.
\end{equation}

So, near $u=0$, we have  from (\ref{a-dos-punts-approx}) and the first equation (\ref{2eqs-aprox}), that is,  
in the Euclidean framework of a pre-cosmology dominated by the $\mu$ term, 
\begin{equation}\label{aEaprox}
a_E \simeq a_{E_{e}}\, \sqrt{\frac{u}{u_e}}, \quad \mu_{E} \simeq \mu (u), 
\end{equation}
$a_{E_e}$ standing for the value of $a_E$ in some ``instant'', $u_e$,  of  the ``energy'' instanton phase.

As explained, this identification of $\mu_E$ with $\mu$ given in the second equation of (\ref{aEaprox}) is only required (by imposing the corresponding extended Darmois condition (\ref{matching-mu})) in the limit $u \to 0$. But, when we go away from $u = 0$, we cannot keep the second equation of (\ref{aEaprox}), since  then, in Eq. (\ref{flu-insta1F+}) the term $-\Lambda/3$ in particular will not stay negligible in front of the term $\kappa \mu/3$ in the same equation, and similarly for the term $-\Lambda/3$ in front of $- \kappa(\mu-3p)$ in Eq. (\ref{flu-insta2F+}). Consequently, $a$ and $a_E$ will not satisfy anymore the same kind of differential equations: function $a$ will satisfy Eqs. (\ref{flu-insta1F}) and  (\ref{flu-insta2F}) with $\epsilon = -1$ while $a_E$ will satisfy Eqs. (\ref{flu-insta1F+}) and  (\ref{flu-insta2F+}) which are not the same pair of equations than (\ref{flu-insta1F}), (\ref{flu-insta2F}). The final conclusion is that, out of $u \to 0$,  $\mu_E$ and $\mu$ are not the same function of $u$.

Nevertheless, we could make the natural guess of keeping for the instanton when $u \neq 0$ the state equation $p_E = - \mu_E/3$ valid for $u \to 0$. In this case, 
Eq. (\ref{sobre-n-orto-E}) keeps giving
\begin{equation}\label{muErad}
\mu_E = \mu_{E_e} \Big( \frac{a_{E_e}}{a_E}\Big)^4, 
\end{equation}
now everywhere, such that $\mu_E$ and $\mu$ are the same kind of function, as functions of $a_E$ and $a$, respectively, although, since we have left $u=0$, $a_E$ and $a$ are no more the same function of $u$. We will use (\ref{muErad}) in the subsection \ref{sec:5b}.

Later on, at the subsection  \ref{sec:5c}, we will consider the case of a general $\mu_E$ function satisfying (\ref{matching-mu}).

As remembered above, for the case of the metric (\ref{LTB-instF}), $K_{ij}$ becomes $K_{ij} = - \frac{1}{2}\partial_u g_{ij}$. In order to have $K_{ij}=0$ for $u \to 0$, we should have, in particular, $\displaystyle{\lim_{u \to 0} \dot{a}= 0}$. But according to (\ref{aLaprox}), this limit does not vanish and becomes in fact $+\infty$. This means that, as announced at the beginning of the present section, just before subsection \ref{sec:4a}, the condition $K_{ij} = 0$ on the junction surface would make any $\Lambda$FLRW universe, closed or not, not quantically creatable, out of the particular case when $\mu = 0$, that is the case considered by Vilenkin \cite{Vilenkin-1983}, who concludes the quantum creatability of the corresponding closed model, the closed de Sitter model.


\section{On the quantum creation of a closed $\Lambda$FLRW universe}
\label{sec:5}

As showed in the precedent section, the instantons corresponding to a closed $\Lambda$FLRW universe are driven by a very ``hot'' phase when $u \to 0$, as a consequence of the imposed ``extended Darmois matching conditons'' (see assumption (b) in Sect. \ref{sec:4}). Could such universe be quantically created from some corresponding instanton? We are going to see that this is possible in the framework we have just designed: more specifically we assume (\ref{disco}) and (\ref{epsilonk}) on the junction surface. However, we will first assume the natural though particular guess $p_E = - \mu_E/3$, $\forall u$, before considering the general case. Because of the Eq. (\ref{sobre-n-orto-E}) this guess is equivalent to assume that $\mu_E \propto a_E^{-4}$.


\subsection{The closed instanton action  $S_E$} 
\label{sec:5a}

In accordance with \cite{Hawking-Ellis},  in an evident notation,  the expression for the Einstein-Hilbert action in the Lorentzian case (including the cosmological constant term) is
\footnote{Notice that we do not include boundary surface terms in the action. In fact, the Einstein field equations (both, for Lorentzian and Euclidean metrics $g_{\alpha\beta}$) follow when the functional action remains stationary under variations of the metric field, $\delta g_{\alpha\beta}$, and its derivatives, $\delta (\partial_\gamma g_{\alpha\beta})$,  which  vanish on the three-boundary of the considered variational four-dimensional domain. For an extensive account of these boundary terms (that have to be included in the action in order to derive the Einstein field equations under arbitrary metric derivatives variations, $\delta(\partial_\gamma g_{\alpha\beta})$) see, for instance, \cite{HaHu,Poisson,Guarnizo-Ca-Te-2010}. \label{sobre-accio}}
%
 %
\begin{equation}\label{EH-accio}
S =  \int [ \frac{1}{2 \kappa} (R - 2 \Lambda) + L] \, \sqrt{\rm{-g}} \, d^4 x. 
\end{equation}

Then, we must consider the extension of the above Einstein-Hilbert action to the instanton case. Actually, following the standard variational procedure (the one displayed, for instance, in \cite{Hawking-Ellis}) it is easy to prove that for an Euclidean metric the action 
\begin{equation}\label{accioE2}
S_{E} =  \int \xi \, \sqrt{\rm{g}_{E}} \, d^4 x, \quad \xi \equiv \frac{1}{2 \kappa} (R_{E} - 2 \Lambda) + L_E, 
\end{equation}
leads to the same expression for the field  equations, that is  $R_{\alpha\beta} - (\frac{R}{2} - \Lambda)  g_{\alpha \beta} = \kappa T_{\alpha\beta}$, for this Euclidean case as well as for the Lorentzian one. Now $\rm{g}_E$ is the instanton metric determinant, ${\rm g}_E \equiv {\rm det}(g_E)_{\alpha\beta}$, that is, the $E$ index refers to the instanton metric, $R_{E}$ being its scalar curvature, and $L_{E}$ its Lagrangian density. We take the global sign of the curvature tensor and the Ricci tensor as showed in the Appendix \ref{ap-B}. Our point of view is that the appropriated instanton Lagrangian $L_{E}$ is the one that is compatible with the above prescription (in our scheme, to derive from an action principle the same form of the Einstein field equations for both Lorentzian and Euclidean metrics). 

In the present work, we only need to calculate $L_{E}$ for a source describing a ``perfect fluid'' instanton. Following a reasoning similar 
to the one displayed in \cite {Hawking-Ellis,Barrow-1993}, according to Eq. (\ref{isotropic-action}) (see Appendix \ref{ap-C}),  the final result is 
\begin{equation}\label{LE}
L_E =   \mu_E.  
\end{equation}

From (\ref{Tmunu2}), we obtain $(T_E)^\alpha_\alpha = \mu_E + 3p_E$ and, taking into account that $L_E = \mu_E$, the expression for $\xi$ in (\ref{accioE2}) becomes
\begin{equation}\label{integrant-accio}
\xi = \frac{1}{2} (\mu_E - 3p_E) +  \frac{\Lambda}{\kappa}, 
\end{equation}
since, using the Einstein field equations,  $R_E = - \kappa (T_E)^\alpha_\alpha  + 4 \Lambda$. Notice that we must use it since, in the semiclassical 
approximation leading to (\ref{proba}), $S_E$ is the instanton classical action, that is, the one calculated upon the classical dynamical trajectories, i.e., upon the suitable solution of the Einstein field equations.

For a general closed FLRW instanton metric, in the coordinates used in (\ref{inst-deSitter-tancat-chi}) we have
\begin{equation}\label{gE}
\sqrt{\rm{g}_{E}} =  a_E^3 \, \sin^2 \chi \, \sin \theta,  \, \, \,   \chi \in [0, 2 \pi], \, \theta \in [0, \pi], 
\end{equation}
and the instanton action (\ref{accioE2}) becomes, with $\xi$ given by (\ref{integrant-accio}),
\begin{equation}\label{SE-mu-gen} 
S_{E}  =  4 \pi^2  \int_{0}^{u_1} \xi \, a_E^3 \, du  = 
4 \pi^2 \int_{a_{E_0}}^{a_{E_1}} \xi \, a_E^3 \, \frac{d a_E}{\dot{a}_E}, 
\end{equation}
where the integration domain, $I$, for the $u$ coordinate is $I = [0, u_1]$ and where 
$a_{E_0} \equiv a_{E}(0)$, $a_{E_1} \equiv a_{E}(u_1)$.


\subsection{The instanton action $S_E$ for a very hot closed $\Lambda$FRW universe} 
\label{sec:5b}

In the present subsection we are going to show the finite character of the instanton  action $S_E$, for the unique  instanton we have associated to a very hot closed $\Lambda$FLRW universe in the above subsection \ref{sec:4c}. This unique instanton comes from our extended Darmois matching conditions, beyond $u=0$  postulating for its state equation $p_E = - \mu_E/3$, the final result being Eq. (\ref{muErad}): $\mu_E = \mu_{E_e} a_{E_e}^4/a_E^4$.

In order to show the above finite character of $S_E$, we will prove first that $\mu_E$ reaches a stationary value $\mu_{E_m}$ for some $a_E$ stationary value, $a_{E_m}$, that is, a value where
\begin{equation}\label{a-dot0}
\dot{a}_E = 0, 
\end{equation}
which becomes necessarily a maximum value since from (\ref{flu-insta2F+}) with $p \equiv p_E = - \mu_E/3$ we always have
\begin{equation}\label{a-Max}
\ddot{a}_E < 0. 
\end{equation}

Then, it is easy to see that, in the present case characterized by (\ref{muErad}), the integrand in (\ref{SE-mu-gen}) is regular for $u=0$, that is, for $a_E \to 0$. 
Notice that for $a_E \to 0$, $\xi \propto a_E^{-4}$, while $a_E/\dot{a}_E$, according to (\ref{flu-insta1F+}), goes like $a_E^2$:  in all a regular value for $\xi a_E^3/\dot{a}_E$. 
Thus, $S_E$ given by the integral (\ref{SE-mu-gen}) has  in the present very hot case a regular integrand in all the integration domain $(a_E = 0, a_E =a_{E_m})$ out of the edge value 
$a_E =a_{E_m}$ where this integrand has a pole because of the denominator $\dot{a}_E$. 

In order to see that this pole does not lead to a divergent value for $S_E$, let us consider the polynome ${\cal P}(1/a_E)$, 
\begin{equation}\label{Pcom}
{\cal P}(1/a_E) \equiv \frac{1}{a_E^4} + \frac{\alpha}{a_E^2} - \frac{\alpha \Lambda}{3}, \, \,  \alpha \equiv \frac{3}{\kappa \mu_{E_m} a_{E_m}^4} >0, 
\end{equation}
that equated to zero is proportional to $\dot{a}_E^2/a_E^2 = 0$ according to (\ref{flu-insta1F+}).

It is easy to see that the fourth degree algebraic equation ${\cal P}(1/a_E)=0$ has two opposite real roots and two pure imaginary ones (let us say $\pm i / r$). 
The positive real root is 
\begin{equation}\label{aM-1}
\frac{1}{a_{E_m}} =  \sqrt{\frac{\alpha} {2} \Big( \sqrt{1 + \frac{4 \Lambda}{3 \alpha}} -1\Big)}, 
\end{equation}
and we have
\begin{equation}\label{Pcom-fac}
{\cal P}(1/a_E) = \Big(\frac{1}{a_E} - \frac{1}{a_{E_m}}\Big) \Big(\frac{1}{a_E} + \frac{1}{a_{E_m}}\Big) \Big(\frac{1}{a_E^2} + \frac{1}{r^2}\Big).
\end{equation}

Then, the contribution, $I_E$,  of the interval $(a_E, a_{E_m})$, when $a_E \to a_{E_m}$, to $S_E$ given by the second integral in Eq. (\ref {SE-mu-gen}), goes like
\begin{eqnarray}\label{Va-com} 
I_{E}  & = &  4 \pi^2 \lim_{a_E \to a_{E_m}} \int_{a_E}^{a_{E_m}} \xi \, a_E^{3} \, \frac{da_E}{\dot{a}_E} \nonumber\\
& \propto &  \int_{a_E}^{a_{E_m}} \, \frac{da_E}{\sqrt{a_{E_m}- a_E}} \propto \sqrt{a_{E_m} - a_E} \to 0
\end{eqnarray}
showing, as it has been claimed, that in the present case of a ``very hot'' instanton,  $S_E$, is finite. Carrying this finite value in (\ref{proba}) we obtain a finite probability of creating a closed $\Lambda$FLRW universe from the corresponding ``very hot'' instanton.

From (\ref{SE-mu-gen}) and (\ref{integrant-accio}), this finite value for $S_E$ is positive and writes out as
\begin{equation}\label{SE-mu-L} 
S_{E}  =  4 \pi^2  \int_{0}^{a_{E_m}} (\mu_E + \frac{\Lambda}{\kappa}) \, a_E^{3} \, \frac{da_E}{\dot{a}_E} 
 = 4\pi^2  \int_{0}^{u_m} (\mu_E + \frac{\Lambda}{\kappa}) \, a_E^{3} \, du,    
\end{equation}
where we have taken into account that now $p_E = - \mu_E/3$ and where $u_m$ is such that $a_E(u_m) = a_{E_m}$.

Let us see that putting here $\mu_E = 0$ we recover the Vilenkin result (\ref{Inst-accio-deS}), as it has to be. 
In fact, for $\mu_E = 0$,   (\ref{SE-mu-L}) becomes
\begin{equation}\label{SE-0-L} 
S_{E}  =  \frac{4 \pi^2  \Lambda}{\kappa}  \int_{0}^{u_1} a_E^{3} \, du, 
\end{equation}
that having in mind (\ref{a-int-dS-knegativa}), setting for the integration limits $a_E(0) = \sqrt{\frac{3}{\Lambda}}$  and $a_E(u_1) = 0$,  gives
\begin{equation}\label{SE-0-cos} 
S_{E} =  \frac{9 \pi}{2 G \Lambda}  \int_{0}^{\pi/2} \cos^3 x  \, dx = \frac{3\pi}{G\Lambda} , 
\end{equation} 
that coincides with (\ref{Inst-accio-deS}). Thus, (\ref{SE-mu-L}) is a generalization to a very hot closed 
$\Lambda$FLRW universe of the $S_E$ action calculated by Vilenkin in the de Sitter case.
%


\subsection{Contrasting the instanton method with other approaches} 
\label{sec:5c}

Notice that the method used in the precedent subsection is not the same that the one qualitatively considered in \cite{Vilenkin-1999}. In this reference, although not explicitly calculated, the $P$ probability in (\ref{proba}), for the case of a hot $\Lambda$FLRW universe, is only considered when there is a classically forbidden region for the evolution of the cosmic expansion factor, $a$. From this region our present universe would emerge by quantum tunneling. Differently to this, our generalization works irrespective of whether this forbidden region exists or not (which depends on whether the maximum of the effective potential is smaller or not than the corresponding ``total'' energy). Our generalization leads to the probability of the quantum creation of a suitable classical universe directly from the corresponding Euclidean solution of the Einstein field equations where there is no time. 

But, why an Euclidean solution of the Einstein field equations should be accepted as the initial state from which creation takes place? As many authors in the field, we consider that the state previous to the universe creation must be a no time state in order that we cannot ask anymore what was before. Another way of getting rid of this question is to introduce a ciclic time (like in \cite{Bervinsky-2007,Bervinsky-KKS-2010}),  or to choose the equation of state so as to generate a pre-tunneling {\em static} configuration 
\cite{AtkaPag-82}, or even to begin from ``nothing'' ($a=0$ with non singular energy density) as in the Vilenkin \cite{Vilenkin-1999} case of a closed de Sitter universe. But, in front of all these proposals, we find that a more convincing view is to move to the Euclidean solutions of the Einstein equations, where the time dimension is substituted by a fourth space dimension. In our opinion, a view worth at least of some initial consideration, as we do in the present paper. 

Notice in any case that, in the present case of a very hot FLRW universe, the cosmic creation probability calculated by us is different from the one qualitatively considered in 
\cite{Vilenkin-1999} (though, in both cases the probability becomes finite). In order to see it, remember that this semiclassical probability, $P$,  is roughly speaking 
$P \propto \exp(-S_E)$,  where in our hot FLRW case $S_E$ is given by Eq. (\ref{SE-mu-L}). On the other hand, in \cite{Vilenkin-1999}, we find for this semiclassical probability 
$P \propto \exp(-2 \int_{a_1}^{a_2}|p(a)| da)$, with $p = - a \dot{a}$ satisfying the equation $p^2 + a^2 - (a^2/a_0)^2 = total \, energy$ (Eqs. (4) and (3) of  \cite{Vilenkin-1999}). A simple checking shows that both expressions for $P$ are clearly different, except in the particular case of $\mu = 0$ (the closed de Sitter universe) in which, as explained before, both coincide.

Thus our approach  and the one from Vilenkin  \cite{Vilenkin-1999} to the calculation of the creation probability of a very hot closed FLRW universe are different, both in the method followed and in the creation probabilities obtained, though both probabilities get a non-vanishing value.

Similarly when we compare our calculated probability with the one obtained in \cite{AtkaPag-82}: to begin with, the respective methods are again different. In \cite{AtkaPag-82}, the equation of the state is chosen so as to reach an ``static'' initial configuration from which the present FLRW universe would appear by quantum tunneling, while in our method there is no tunneling and the Euclidean action, $S_E$, is calculated from the corresponding Euclidean solution of the Einstein field equations without performing any Wick rotation. Furthermore, the $S_E$ value obtained by us and the one in  \cite{AtkaPag-82} are qualitatively different: we only have to notice that in the basic cosmic equation of \cite{AtkaPag-82}, Eq. (27), and in the corresponding Eq. (\ref{flu-insta1F+}) of the present paper,  the terms in $k$ have opposite sign. Then, if these two equations are different, the corresponding $S_E$ values will be different too. This difference can be retraced to the fact that, in the present case of a very hot FLRW universe, the ansatz $t \to - i \, t$ used in \cite{AtkaPag-82} to obtain $S_E$ and our method of selecting the suitable Euclidean solution of the Einstein equations are not equivalent.


\subsection{The finite character of the closed instanton action for a general $\mu_E$ function} 
\label{sec:5d}

Let us  come back to (\ref{SE-mu-gen}) the general expression for the closed instanton action:
\begin{equation}\label{SE-mu-gen-2} 
S_{E}  = 4 \pi^2 \int_{a_{E_0}}^{a_{E_1}} \xi \, a_E^3 \, \frac{d a_E}{\dot{a}_E} 
\end{equation}
where $\xi \equiv \frac{1}{2} (\mu_E - 3 p_E) + \frac{\Lambda}{\kappa}$, and where $\mu_E$ is now a general function of $u$ satisfying the limit condition (\ref{matching-mu}).

First, provided that we have $|p_E| \leq |\mu_E|$, the above integrand has no singularity for $a_E \to 0$ since, in accordance with Sec. \ref{sec:4c}, for $a_E \to 0$, we have 
again $\mu_E \propto a_E^{-4}$, while the factor $a_E^3/\dot{a}_E$ goes like $a_{E}^4$. Contrarily to this, we have a singularity for $\dot{a}_E = 0$. 
Then imagine that $a_{E_m}$ is the minimal positive value of $a_E$ such that $\dot{a}_E = 0$. In this case, a sufficient  condition to still have a finite value for 
$S_E$ is that, for $a_E \to a_{E_m}$, $a_E$ behaves such that  $\dot{a}_E / a_E \propto (a_{E_m} - a_E)^{n}$, with $0 \leq n<1$, that generalizes the condition for having a finite value of $S_E$ considered in the precedent subsection \ref{sec:5b}.

Thus, the guess $p_E = - \mu_E/3$ is by no means the only state equation for a hot closed $\Lambda$FLRW universe, whose instanton action $S_E$ is finite making this universe quantically creatable.


\section{Summary and concluding remarks}
\label{sec:6}

In the present paper we have extended the seminal work by Vilenkin in \cite{Vilenkin-1982,Vilenkin-1983,Vilenkin-1984,Vilenkin-1985,Vilenkin-1999} where the semiclassical 
probability $P$ of creating from ``nothing'' a closed de Sitter universe is calculated. For Vilenkin, $P$ is the probability of the corresponding quantum tunneling event, leading to the creation event of this universe from ``nothing'', i. e., from the appropriate instanton (Euclidean solution).

Without relying on cosmic quantum tunneling, we have completed this work by stating a general method  to find the instantons corresponding to any universe belonging to a large family of Lorentzian metrics (including the LTB metrics).  In particular, we have calculated the quantum creation probability of the FLRW metrics  in the semiclassical approximation. This general method includes the particular case used in \cite{Vilenkin-1982,Vilenkin-1983,Vilenkin-1984,Vilenkin-1985}, finding the instanton from which the closed de Sitter  universe would be created. 

When applied to the case of a general $\Lambda$FLRW universe, this method allows us to recover the result from \cite{AtkaPag-82}, according to which the only FLRW universe which becomes quantically creatable is the closed one. However, the method used in \cite{AtkaPag-82} is different from ours: in \cite{AtkaPag-82} the closed universe is created by quantum tunneling from some initial static space-time configuration.

Prior to the application of our general method to assess the possibility of the quantum creation of a general $\Lambda$FLRW universe, we have cast the Einstein field equations in a simplified canonical form valid, both, for a Lorentzian metric, and for an instanton, in the particular case of (inhomogeneous) spherical symmetry, with a stress-energy tensor corresponding to a geodesic perfect  fluid, with cosmological constant. The general solution of these canonical equations has been given explicitly in the special case of the LTB universes. This explicit solution is very well known, but we give it here for completeness.

Moreover, leaving the equation of state, $p = \mu/3$, used in the subsection \ref{sec:4c}, we have provided some general conditions for finding a finite probability of creating a closed $\Lambda$FLRW universe. But, what about the two remaining cases, the open non flat and the flat ones? Following a similar approach to that of Sect. \ref{sec:5}, but for the $\Lambda$FLRW curvature index $k_L$ equal to $-1$ or to zero, it is easy to see that the action $S_E$ of the corresponding instanton diverges positively. This indicates that  these two $\Lambda$FLRW universes, the open non flat and the flat ones, are not quantically creatable in the framework of our general method. In order to see this notice that, instead of the instanton metric (\ref{inst-deSitter-tancat-chi}), we must use the following one
\begin{equation}\label{inst-deSitter-obert-chi}
ds^2_E=   du^2+ a^2_E(u) ( d \chi^2 + \sinh^2 \chi \,  d\sigma^2),  
\end{equation}
in the open non flat case, or 
\begin{equation}\label{inst-deSitter-pla-chi}
ds^2_E=   du^2+ a^2_E(u) ( d \chi^2 + \chi^2 \,  d\sigma^2),  
\end{equation}
in the flat case. In both cases $\chi$ takes values from zero to infinite. This leads directly to a positive infinite value of $S_E$, since in the calculation of $S_E$,  given by the multiple definite integral (\ref{accioE2}),  the one variable definite integral, $\int_0 ^\infty \sinh^2 \chi \, d \chi$ or $\int_0 ^\infty \chi^2 \, d \chi$, respectively, appears which are both obviously divergent.

However, this instanton action $S_E$ can be rendered finite for the considered flat and open FLRW universes if some non trivial topologies for their corresponding 3-spaces are considered. In this way, the flat and open FLRW universes would become quantically creatable \cite{Zeldovich-Starobinsky-84,Gurzadyan-Kacharyan-89}. 

In summary, in our method, the closed $\Lambda$FLRW universe becomes creatable, but not the open non flat or the flat ones with a trivial topology for their corresponding 3-spaces. 
On the other hand, in \cite{Lapiedra-Morales-2012,Lapiedra-Morales-2013}, where we extended some previous results \cite{Ferrando,Lapiedra-Saez}, we concluded that what we call the {\em intrinsic} space-time 4-momenta vanish in the particular case of a closed and  a flat FLRW universes, while the {\em intrinsic} energy, in particular, diverges in the open case. But in \cite{Lapiedra-Morales-2012}, confirming a previous result in \cite{Lapiedra-Saez}, the flat FLRW model, when perturbed in the framework of the standard inflation, has finally infinite {\em intrinsic} energy. Thus we could consider the vanishing of 4-momenta of this flat model as an unstable, and so unphysical, result. So, we could consider both, the open non flat, and the flat, FLRW models, as having a non finite energy. In any case, we could never be able to confirm observationally whether we live in a flat FLRW universe: an apparent flat FLRW universe could actually be an open or closed one with a large enough curvature radius .

All this seems in accordance with the idea announced in \cite{Fomin-73,Tryon,Guth-llibre,Linde-1984a,Linde-1984b} according to which the energy of our universe would have to vanish for the universe to be created as a vacuum quantum fluctuation. Sometimes this idea has been discarded from the very beginning by arguing that the energy of a space-time is only properly defined when the space-time is asymptotically Minkowskian (see for example \cite{AtkaPag-82}) and so the energy definition would become inapplicable to the particular $\Lambda$FLRW universe case. However, if we start from an energy complex (such as the Weinberg one \cite{Weinberg} selected in the \cite{Lapiedra-Morales-2012,Lapiedra-Morales-2013,Ferrando,Lapiedra-Saez} papers) to define the energy and, in general, the linear and angular 4-momenta of the considered space-time, one can be easily convinced that these two 4-momenta have a precise meaning provided that  their 3-volume integral expressions converge, irrespective of whether the space-time is asymptotically Minkowskian or not. Actually, the meaning of the corresponding energy, for example, is that of being a global quantity whose different non additive energy components, gravitational or not, enter into their mutual balance. This generalizes to General Relativity the similar balance we find in elementary physics where, in order to define the global energy of a system, we define successive different energies (potential, electromagnetic, ...) entering into balance with the kinetic one inside this global  energy. Similarly for the rest of the 4-momenta components in General Relativity.

The mathematical translation of this balance to General Relativity is the vanishing of the ordinary 4-divergence of the corresponding energy-momentum complex irrespective of whether the space-time is asymptotically Minkowskian or not. In fact the real problem of these two 4-momenta  is that, even when they exist, they are dramatically dependent on the coordinate system used. This is the reason why in  \cite{Lapiedra-Morales-2012,Lapiedra-Morales-2013} we have defined {\em intrinsic} coordinates for a universe, and  then selected the corresponding {\em intrinsic} 4-momenta as the ones that according to \cite{Fomin-73,Tryon,Guth-llibre} should vanish for the corresponding universe to be quantically creatable.

In summary, our results in the present paper seem to be in accordance with such a view since the FLRW universe which becomes quantically creatable, the closed one,  is just the one with vanishing {\em intrinsic} 4-momenta.

Finally, as most authors in the field, we assume that these results, obtained in the semiclassical approximation, would remain essentially the same when going to upper 
orders in $\hbar$. All the same, instead of going to this upper level it would seem more interesting to follow \cite{Bervinsky-2007,Bervinsky-KKS-2010} and to consider the case of the standard inflaton non minimally coupled to curvature. This procedure would generalize the lowest case of a closed de Sitter model, with this generalization  performed in our approach  of the instantons as mere Euclidean solution of the Einstein equations. However, this would be beyond the scope of the present paper and would deserve future work.

\vspace{1cm}


{\bf Acknowledgements} This work was supported by the Spanish 
``Ministerio de  Econom\'{\i}a y Competitividad'', MICINN-FEDER project FIS2012-33582. Useful discussions with J. Navarro-Salas are pleasingly recognized.


\appendix

\section{Isotropic Instanton and  ``conservation equations'', $\nabla \cdot{T} = 0$}
\label{ap-A}

Let $V_4$ be a four-dimensional manifold and $g$ a metric on $V_4$ whose signature is $(\epsilon, +, +,  +)$, 
with $\epsilon = -1$ (Lorentzian case) or $\epsilon = 1$ (Euclidean case) and 
let us consider an isotropic stress-energy tensor $T$ (see \cite{Ellis-1992})%
\begin{equation}\label{A-fluper-instanto}
T = \mu \, n \otimes n + p \, \gamma,
\end{equation}
where  $\gamma \equiv  g - \epsilon \, n \otimes n$,  \,  $n^2 \equiv g(n, n) = \epsilon$, and
\begin{equation}\label{A-autovalors}
\lambda_s \equiv \epsilon \, \mu,  \qquad  \lambda_t \equiv p, 
\end{equation}
with $\lambda_s$ ($\lambda_t$) the simple  (triple) eigenvalue of $T$,  
\begin{equation}\label{A-autovectors}
T(n) = \epsilon \,  \mu \,  n, \quad T(v) = p \, v 
\end{equation}
for any vector $v$ ortogonal to $n$, $g(n,v) = 0$. Then 
\begin{equation}\label{A-fluper-instanto-2}
T = \epsilon \,  (\lambda_s - \lambda_t) \, n \otimes n + \lambda_t \, g = (\mu - \epsilon \, p) \, n \otimes n + p \, g,  
\end{equation}
that, in the Lorentzian case, is the expression for the  stress-energy tensor  of a perfect fluid, 
\begin{equation}\label{A-fluper}
T_{L} = (\mu + p) \, n \otimes n + p \, g,
\end{equation}
and, in the Euclidean case,  it becomes 
\begin{equation}\label{A-instanto}
T_{E} = (\mu - p) \, n \otimes n + p \, g, 
\end{equation}
describing the ``stresses'' associated to an instanton field.

For a conserved $T$, $\nabla \cdot T =0$, the divergence of (\ref{A-fluper-instanto}) gives 
\begin{eqnarray}\label{divT}
\nabla_\alpha T^{\alpha \beta} & = & u^\alpha \partial_\alpha (\mu - \epsilon p)  \, n^\beta \nonumber \\ & + & (\mu - \epsilon p) [\theta  \, n^\beta + a^\beta] + (\partial_\rho p) \, g^{\rho\beta} = 0, 
\end{eqnarray}
with $\theta \equiv \nabla_\alpha n^\alpha$, $a^\beta \equiv n^\alpha \nabla_\alpha n^\beta$, whose decomposition in parts parallel and orthogonal to $n^\alpha$ leads to
\begin{equation}\label{energy-inst}
\dot{\mu} = - (\mu - \epsilon p) \, \theta, 
\end{equation}
and 
\begin{equation}\label{Euler-inst}
(\mu - \epsilon p) \, a_\beta  = - \partial_\beta p + \epsilon \, \dot{p} \, n_\beta, 
\end{equation}
respectively, where the dot denotes derivation along $n^\alpha$.


\section{Spherically symmetric Einstein equations for both, instantons and Lorentzian solutions with geodesic perfect flows}
\label{ap-B}

The material in this appendix can be found in most relativity textbooks  that mainly deal with the Lorentzian case (see for example \cite{PlebKra}). The case of a general diagonal metric is considered in  \cite{Rindler}. This material is included here for sake of completeness 
and in order to precise the sign conventions we use. 

To begin with,  we take for the Riemann tensor definition of $(V_4, g)$
\begin{equation}\label{Riemann}
R^\alpha_{\,\, \beta\mu\nu} = \partial_\mu \Gamma^\alpha_{\nu\beta} - \partial_\nu \Gamma^\alpha_{\mu\beta} + 
\Gamma^\alpha_{\mu\lambda} \Gamma^\lambda_{\nu\beta} - \Gamma^\alpha_{\nu\lambda} \Gamma^\lambda_{\mu\beta}, 
\end{equation}
and for the Ricci tensor
\begin{equation}\label{Ricci}
R_{\mu\nu} = R^\alpha_{\,\,\mu\alpha\nu}.
\end{equation}

If the metric $V_4$ is spherically symmetric, in a Gauss coordinate system, $\{u, \rho, \theta, \phi\}$,  adapted to the symmetry  the line element is written as:
\begin{equation}\label{Ap-metric}
ds^2 = \epsilon \,du^2  + B(u, \rho)\,d\rho^2 + C(u, \rho)\, d\sigma^2
\end{equation}
where $d \sigma^2 = d\theta^2 + \sin^2 \hspace{-0.8mm} \theta \, d\phi^2$ stands for  the metric on the unit two-sphere.  According to the sign conventions (\ref{Riemann}) and (\ref{Ricci}), the essential components of the Ricci tensor in these coordinates are given by:
\begin{eqnarray}\label{Riicci-Gauss}
R_{uu} & = & - \frac{\ddot B}{2B} -  \frac{\ddot C}{C} +  \frac{\dot B^2}{4B^2} +  \frac{\dot C^2}{2C^2}, \\
R_{\rho\rho} & = & - \frac{\epsilon}{2} \Big(\ddot B -  \frac{\dot B^2}{2B} +  \frac{\dot B \dot C}{C}\Big) +  \frac{1}{C} \Big(- C'' + \frac{C'^2}{2C} + \frac{B' C'}{2 B} \Big), \\
R_{\theta\theta} & = & - \frac{\epsilon}{2} \Big(\ddot C + \frac{\dot{B} \dot{C}}{2B}\Big) - \frac{1}{2B}\Big(C'' - \frac{B' C'}{2B} \Big) +1,\\  
R_{\phi\phi} & = & R_{\theta\theta} \sin^2 \theta, \\
R_{u\rho} & = & \frac{1}{C} \Big( - \dot{C}' + \frac{\dot{C} C'}{2C} + \frac{\dot{B} C'}{2B}\Big),  
\end{eqnarray}
the remaining components being identically zero by virtue of the assumed symmetry. 

The Einstein field equations 
\begin{equation}\label{A-Einstein}
Ric(g) - \frac{R}{2} g + \Lambda g = \kappa T,  \quad  \kappa \equiv 8 \pi G, 
\end{equation}
are written in equivalent form as
\begin{equation}\label{A-Einstein-equiv}
Ric(g) = \kappa T + (\Lambda- \frac{\kappa}{2} {\rm tr} T) \, g. 
\end{equation}

Then for an isotropic source $T$ given by (\ref{A-fluper-instanto}),  Eq. (\ref{A-Einstein-equiv}) becomes
\begin{equation}\label{A-Einstein-flu-in}
Ric(g) = \epsilon[\Lambda + \frac{\kappa}{2} (\lambda_s - 3 \lambda_t)] \, n \otimes n + [\Lambda -  \frac{\kappa}{2} (\lambda_s - \lambda_t)] \, \gamma. 
\end{equation}
When the field $n$ is geodesic, that is, when $\partial_\rho p =0$, 
adapting Gauss coordinates  $(u, \rho, \theta, \phi)$ so that $n= \epsilon \, du = (\epsilon, 0, 0, 0)$, the  Einstein equations are:
\begin{equation}\label{A-Einstein-1}
 - \frac{\ddot B}{2B} -  \frac{\ddot C}{C} +  \frac{\dot B^2}{4B^2} +  \frac{\dot C^2}{2C^2}  =  \epsilon \,  [ \Lambda + \frac{\kappa}{2} (\lambda_s - 3 \lambda_t)],  
\end{equation} 
\begin{equation}\label{A-Einstein-2}
- \frac{\epsilon}{2} \Big(\ddot B -  \frac{\dot B^2}{2B} + \frac{\dot B \dot C}{C}\Big) + \frac{1}{C} \Big(- C'' + \frac{C'^2}{2C} + \frac{B'C'}{2 B} \Big)  
=  [ \Lambda - \frac{\kappa}{2} (\lambda_s + \lambda_t)] B, 
\end{equation}
\begin{equation}\label{A-Einstein-3}
 - \frac{\epsilon}{2} \Big(\ddot C + \frac{\dot{B} \dot{C}}{2B}\Big) - \frac{1}{2B}\Big(C'' - \frac{B' C'}{2B} \Big) +1  = [ \Lambda - \frac{\kappa}{2} (\lambda_s + \lambda_t)] C, 
\end{equation}
\begin{equation}\label{A-Einstein-4}
 - \dot{C}' + \frac{\dot{C} C'}{2C} + \frac{\dot{B} C'}{2B} = 0.
\end{equation}

The integration of these equations involves two separed cases. The case $C'=0$ leads to a generalized family of Datt metrics with its associated instanton family, and will be analized elsewhere. Here, let us consider the generic case $C' \neq 0$. Then (\ref{A-Einstein-4}) is written as:
\begin{equation}\label{A-Einstein-4b}
\frac{\partial}{\partial u}\Big(\ln \frac{C'^{2}}{CB}\Big) = 0,
\end{equation}
and then $\chi \equiv C'^{2}/ CB$ does not depend on the $u$ coordinate, i.e., 
\begin{equation}\label{A-functionB}
B(u, \rho) =  \frac{C'^{2}}{C \chi(\rho)} = \frac{4 A'^{2}}{\chi(\rho)}, 
\end{equation}
where we have put $C \equiv A^2$. Defining $\chi (\rho) \equiv 4 (1 + \epsilon K(\rho))$, the metric (\ref{Ap-metric}) becomes: 
\begin{equation}\label{A-LTBmetric}
ds^2 = \epsilon du^2 + \frac{A'^{2}}{1 + \epsilon K(\rho)} d\rho^2 + A^2 d \sigma^2.
\end{equation}

The following linear combination
\begin{equation}\label{Eq-combi}
\displaystyle{{\rm Eq. \, (\ref{A-Einstein-1})} - \frac{\epsilon}{B} \, {\rm Eq. \, (\ref{A-Einstein-2})}  + 2 \, \frac{\epsilon}{C} \,  {\rm Eq. \, (\ref{A-Einstein-3})}}
\end{equation}
leads to  
\begin{equation}\label{EqC}
-2 \frac{\ddot C}{C} + \frac{\dot{C}^2}{2 C^2} + \frac{\epsilon}{C}(2 - \frac{C'^2}{2 B C}) = 2 \epsilon \, (\Lambda - \kappa \,  \lambda_t)
\end{equation}
and taking into account (\ref{A-functionB}) with $C=A^2$ we obtain the following equation for $A$:
\begin{equation}\label{A equation}
\dot{A}^2+2A\ddot{A} + K + \epsilon \, \Lambda \, A^2 = \epsilon \, \kappa \, \lambda_t  \, A^2. 
\end{equation}
By substituting (\ref{A equation}) in (\ref{A-Einstein-1}), we arrive to
\begin{equation}\label{A equation 2}
2 \frac{\ddot{A}}{A}+\frac{\ddot{A'}}{A'} + \, \epsilon \,  \Lambda =- \frac{\kappa}{2} \, \epsilon \, (\lambda_s -3 \lambda_t).    
\end{equation}
Then,  the remaining Einstein equations (\ref{A-Einstein-2}) and  (\ref{A-Einstein-3}) are identically satisfied by substitution of  (\ref{A equation}) and (\ref{A equation 2}) in them.

Therefore, the Einstein equations for the, let us say, generalized $\Lambda$LTB family of metrics (\ref{A-LTBmetric}) (now the pressure $p$ is homogeneous, $p = p(u)$ and can be different from zero) reduce to (\ref{A equation}) and (\ref{A equation 2}), or equivalently, to the following equations:
\begin{equation} \label{A equation 3}
3 \,  \frac{\dot{A}^2}{A^2} + 2 \Big(\frac{\ddot{A}}{A} - \frac{\ddot{A'}}{A'}\Big) + 3 \,  \frac{K}{A^2} + \epsilon \,  \Lambda = \kappa  \, \epsilon \, \lambda_s   
\end{equation}
and
\begin{equation} \label{A equation 4}
2 \,  \frac{\ddot{A}}{A} + \frac{\dot{A}^2}{A^2} + \frac{K}{A^2} + \epsilon \,  \Lambda = \epsilon \, \kappa \,  \lambda_t.   
\end{equation}

In the case $p= \Lambda = 0$, we recover the LTB-family and the associated instanton, which are both governed by the following equations:
\begin{equation}\label{AE1}
\dot{A}^2+2A\ddot{A} + K=0, 
\end{equation}
\begin{equation}\label{AE2}
2 \frac{\ddot{A}}{A}+\frac{\ddot{A'}}{A'}=-4\pi \mu.  
\end{equation}
that do not depend on $\epsilon$.  Nevertheless, despite that the above pair of equations, (\ref{AE1}) and  (\ref{AE2}),  are signature independent, 
the corresponding metric components, $g_{\rho\rho} = A'^{2}/(1 + \epsilon K(\rho))$ depend on the $\epsilon$ sign.

From (\ref{AE1}), it results that $A^2 \ddot{A}$ does not depend on $u$. Defining $M(\rho) \equiv  - A^2 \ddot{A}$, Eqs. (\ref{AE1}) and (\ref{AE2}) are respectively written,   
\begin{equation}\label{A-Apunt2}
\dot{A}^2 = \frac{2}{A} M(\rho) - K(\rho), 
\end{equation}
\begin{equation}\label{A-muLTB}
4\pi \mu = \frac{M'(\rho)}{A^2 A'}.    
\end{equation}
Integration of (\ref{A-Apunt2}) leads to (\ref{kp})-(\ref{kn}).


\section{``Perfect fluid'' instanton Lagrangian}
\label{ap-C}

In this Appendix, we detail the steps allowing to determine the instanton Lagrangian  $L_E = \mu_E$ whose stress-energy tensor is the one considered in Appendix \ref{ap-A}, that is 
(\ref{A-instanto}). We follow closely the method developed in references \cite{Hawking-Ellis,Barrow-1993} for action functionals of matter fields, starting from the expression 
\begin{equation}\label{Tvariacional}
T^{\mu\nu} =  \frac{2}{\sqrt{|\rm{g}|}} \, \frac{\partial(L \sqrt{|\rm{g}|})}{\partial g_{\mu\nu}}  
\end{equation}
where ${\rm g}$ stands for the metric determinant, ${\rm g} \equiv {\rm det}g_{\alpha\beta}$. Let us consider the Lagrangian $L=f(\omega, \xi)$ and its associated action $S = \int f  \sqrt{\epsilon \rm{g}} \,  d^4x$. Substituting this $L$ in (\ref{Tvariacional})
we will determine the function $f(\omega, \xi)$ which leads to the general expression of an isotropic tensor given by (\ref{A-fluper-instanto}) .
We assume that $\omega$ is alike a thermodynamic  variable that has associated a ``conserved flux of particles'', say $N$:
\begin{equation}\label{N-corrent}
N = \omega \, n, \qquad n^2 = \epsilon.
\end{equation}
In addition, we also assume that the $\xi$ variable is constant  on each integral curve of the vector field $N$, that is, 
$n^\alpha \partial_\alpha \xi = \dot \xi = 0$ (``isoentropic flux'') and then, that $f$ only depends on $\omega$, $f = f(\omega)$ (``barotropic instanton'' for $\epsilon = +1$). 
Then, developing  (\ref{Tvariacional}) we obtain %
\begin{equation}\label{Tvariacional2}
T^{\mu\nu} = 2 \, \frac{\partial f}{\partial g_{\mu\nu}} + \frac{f}{\rm{g}} \, \frac{\partial \rm{g}}{\partial g_{\mu\nu}} = 
 2 \,  f'  \, \frac{\partial \omega}{\partial g_{\mu\nu}} + f g^{\mu\nu}
\end{equation}
where $f' \equiv df/d\omega$ and we have considered the relation $g^{\mu\nu} = \frac{1}{\rm{g}} \frac{\partial \rm{g}}{\partial g_{\mu\nu}}$.

On the other hand, from (\ref{N-corrent}),  
\begin{equation}\label{omega2}
\omega^2 = \epsilon \, g_{\alpha \beta} \, N^\alpha N^\beta = \frac{g_{\alpha\beta}}{{\rm g}} (\sqrt{\epsilon {\rm g}} N^\alpha) (\sqrt{\epsilon {\rm g}} N^{\beta}), 
\end{equation}
whose variation with respect to $g_{\mu\nu}$, the flow lines being given, leads to 
\begin{eqnarray}\label{var-omega-2}
2 \omega \, \frac{\partial \omega}{\partial g_{\mu\nu}} =  \Big[\frac{\partial}{\partial g_{\mu\nu}}\, \Big( \frac{g_{\alpha \beta}}{{\rm g}} \Big)\Big] \,  \epsilon \, {\rm g} \, N^\alpha N^\beta =  \nonumber \\ \epsilon \, {\rm g}  \, \omega^2 \, \Big( \frac{1}{\rm g} \delta^\mu_\alpha \delta^\nu_\beta \, - \, \frac{1}{{\rm g}^2} \frac{\partial{\rm g}}{\partial g_{\mu\nu}} \, g_{\alpha \beta }\Big) n^\alpha \, n^\beta
=  \omega^2  ( \epsilon \, n^\mu n^\nu - g^{\mu\nu}), 
\end{eqnarray}
because, given the integral curves,  the ``conserved current'' $\sqrt{{\epsilon \rm g}} N^\alpha$ is uniquely determined on a flow line in terms of its initial value at some point on the same flow line. Then,  this ``conserved current'' remains unchanged for variations of $g_{\mu\nu}$ that vanish on the boundary made out from the points where we fix these initial values.  

Then, substituting in (\ref{Tvariacional2}),  we have:
\begin{equation}\label{Tvariacional3}
T^{\mu\nu} =  
 \omega \,  f'    ( \epsilon \, n^\mu n^\nu - g^{\mu\nu}) +  f  g^{\mu\nu} = 
 \epsilon \, \omega \,  f'  n^\mu n^\nu + ( f - \omega \,  f' ) g^{\mu \nu}, 
\end{equation}
which is an isotropic 2-tensor $T$ of the form (\ref{A-fluper-instanto-2}), with 
\begin{equation}\label{fp}
f = \lambda_s = \epsilon \, \mu,  \qquad    \lambda_t = p = f - \omega \, f'.
\end{equation}

In all, the functional action we are looking for an isotropic stress tensor is given by 
\begin{equation}\label{isotropic-action}
S = \epsilon \int \mu   \, \sqrt{\epsilon \, \rm{g}} \,  d^4x, 
\end{equation} 
with $\epsilon = +1$ and $L_E = \mu_E$ for the instanton case. 

Incidentally note that, in particular, the above expression for the Lagrangian density in (\ref{isotropic-action}), $L = \epsilon \mu$, 
allows us to deduce the $-\Lambda/\kappa$ term appearing in the instanton action $S_E$ (see Eq. (\ref{accioE2})).  In the Einstein field equations, the $\Lambda g_{\alpha \beta}$ term at the  left  member corresponds to a source term $T = - (\Lambda/\kappa) g$ at the right member. Thus, according to (\ref{A-fluper-instanto-2}), this tensor has $\epsilon \mu = p = - \Lambda/\kappa$, and then, it follows from the variations of the Lagrangian density  $L = \epsilon \mu = - \Lambda/\kappa$, with respect to the metric (in both the Lorentzian or the  
Euclidean cases).



\bibliography{apssamp}

\end{document}